\documentclass[conference]{IEEEtran}
\usepackage{amsmath,amsfonts}
\usepackage{graphicx,amssymb}
\usepackage{algorithmic,algorithm}
\usepackage{amsthm,dsfont}
\usepackage{subfigure,url}
\usepackage[noadjust]{cite}
\usepackage{xcolor}
\usepackage{colortbl,booktabs}

\theoremstyle{plain}

\newtheorem{thm}{Theorem}

\theoremstyle{remark}

\newcounter{longequ}[longequ]
\addtocounter{longequ}{11}

\allowdisplaybreaks

\IEEEoverridecommandlockouts
\begin{document}
%
\title{Neural Calibration for Scalable Beamforming in FDD Massive MIMO with Implicit Channel Estimation
}
\author{\IEEEauthorblockN{Yifan Ma\IEEEauthorrefmark{1},
		Yifei Shen\IEEEauthorrefmark{1},  Xianghao Yu\IEEEauthorrefmark{1}, Jun Zhang\IEEEauthorrefmark{2}, S.H. Song\IEEEauthorrefmark{1}\IEEEauthorrefmark{3}, and Khaled B. Letaief\IEEEauthorrefmark{1}}\\
	\IEEEauthorblockA{\IEEEauthorrefmark{1}Dept. of ECE, The Hong Kong University of Science and Technology, Hong Kong\\
	\IEEEauthorrefmark{2}Dept. of EIE, The Hong Kong Polytechnic University, Hong Kong\\
	\IEEEauthorrefmark{3}Division of ISD, The Hong Kong University of Science and Technology, Hong Kong\\
	Email: \IEEEauthorrefmark{1}\{ymabj, yshenaw, eexyu, eeshsong, eekhaled\}@ust.hk, \IEEEauthorrefmark{2}jun-eie.zhang@polyu.edu.hk}}

\maketitle

\begin{abstract}
Channel estimation and beamforming play critical roles in frequency-division duplexing (FDD) massive multiple-input multiple-output (MIMO) systems.
However, these two modules have been treated as two stand-alone components, which makes it difficult to achieve a global system optimality.
In this paper, we propose a deep learning-based approach that directly optimizes the beamformers at the base station according to the received uplink pilots, thereby, bypassing the explicit channel estimation.
Different from the existing fully data-driven approach where all the modules are replaced by deep neural networks (DNNs), a \emph{neural calibration} method is proposed to improve the scalability of the end-to-end design.
In particular, the backbone of conventional time-efficient algorithms, i.e., the least-squares (LS) channel estimator and the zero-forcing (ZF) beamformer, is preserved and DNNs are leveraged to calibrate their inputs for better performance.
The permutation equivariance property of the formulated resource allocation problem is then identified to design a low-complexity neural network architecture.
Simulation results will show the superiority of the proposed neural calibration method over benchmark schemes in terms of both the spectral efficiency and scalability in large-scale wireless networks.
\end{abstract}

\IEEEpeerreviewmaketitle
\section{Introduction}
Massive multiple-input multiple-output (MIMO) is a disruptive technology in wireless systems that can significantly improve wireless system performance, including system capacity \cite{Marzetta10MIMO, Lu14MIMO}. A prevalent operation mode of practical massive MIMO systems is frequency-division duplexing (FDD) which enjoys a lower cost and greater coverage than time-division duplexing (TDD) \cite{Qualcomm}. 
However, the downlink channel state information (CSI) acquisition is a challenging task in FDD massive MIMO systems due to the weak reciprocity between the uplink and downlink channels. The feedback of the downlink CSI causes prohibitive delay and overhead, which is proportional to the number of users.
Meanwhile, conventional beamforming approaches typically employ iterative algorithms, where computationally-intensive procedures are involved in each iteration, leading to a high computational complexity. These two factors make it difficult to support a large number of users in FDD massive MIMO systems.

To overcome the huge estimation overhead, deep learning-based downlink channel prediction for FDD systems has been recently studied \cite{Alrabeiah19FDD}. Given that the uplink and downlink channels share the same physical propagation environment, neural networks were utilized to capture the intricate mapping from the uplink channels to the downlink channels. On the other hand, learning-based beamforming optimization has been investigated in \cite{Lin20Beamform, Shen21Graph} to improve the computational efficiency. This is achieved by replacing conventional iterative beamforming algorithms with deep learning techniques based on multi-layer perceptron (MLP) \cite{Lin20Beamform} or graph neural networks (GNNs) \cite{Shen21Graph}. 
Although channel estimation and beamforming can be performed in real-time via the techniques developed in \cite{Alrabeiah19FDD, Lin20Beamform, Shen21Graph}, separately considering these two modules in a block-by-block manner makes it difficult to achieve global system optimality \cite{Ye20E2E}. 

To address this problem, an end-to-end design has recently been proposed. 
With the powerful learning capabilities of deep neural networks (DNNs), it is possible to directly learn the mapping from the received pilots to the target beamforming vectors without the need of explicit channel estimation \cite{sohrabi2020deep, Jiang20Implicit}. The authors of \cite{sohrabi2020deep} proposed a deep learning approach to jointly design the CSI feedback and downlink beamforming in FDD massive MIMO systems by training a series of MLPs at the users and the BS. A similar approach was extended to the newly-emerged intelligent reflecting surface (IRS)-assisted wireless systems \cite{Jiang20Implicit}, where the passive and active beamforming vectors at the IRSs and BS are learned based on the received pilots, respectively. However, one key limitation of these methods is that they are \emph{not scalable} to large-scale wireless networks. In particular, existing end-to-end approaches typically resort to fully data-driven methods \cite{sohrabi2020deep, Jiang20Implicit}, where conventional signal processing modules are treated as a black-box and replaced by standard neural network architectures. Since these architectures are not customized to specific design problems in wireless systems, the underlying high-dimensional mapping is difficult to learn in large-scale wireless networks. Therefore, the performance of existing end-to-end designs deteriorates dramatically when the network size increases.

In this paper, we develop a scalable deep learning-based method to perform an end-to-end design for FDD massive MIMO systems. Specifically, we jointly consider the design of uplink transmit pilots, implicit channel estimation, and downlink beamforming. Instead of completely replacing traditional handcrafted solutions with DNNs \cite{sohrabi2020deep, Jiang20Implicit}, we preserve the backbones of two classic time-efficient algorithms, i.e., the least-squares (LS) estimator and the zero-forcing (ZF) precoder, and use neural networks to calibrate their inputs for a better performance. 
The permutation equivariance property of the formulated resource allocation problem is then leveraged to design a low-complexity neural network architecture, where one lightweight MLP is reused for different users. It is shown that by exploiting the domain knowledge inherent in the problem-specific handcrafted solutions and the intrinsic equivariance property introduced by the wireless network topology, the resulting end-to-end design provides high scalability in large-scale wireless networks. Extensive simulation results will demonstrate that the proposed method outperforms both the conventional block-by-block solutions and the state-of-the-art end-to-end design. More importantly, the scalability of the proposed neural calibration design is verified in FDD massive MIMO systems.

\begin{table*}[t]	
\selectfont  
\centering
\newcommand{\tabincell}[2]{\begin{tabular}{@{}#1@{}}#2\end{tabular}}
\caption{Typical Methodologies for The Conventional Block-by-Block Paradigm.} 
\begin{tabular}{|c|c|c|c|}
\hline
  & \tabincell{c}{Block 1: \\ Uplink channel estimation} & \tabincell{c}{Block 2: \\ Downlink channel prediction} & \tabincell{c}{Block 3: \\ Downlink beamforming} \\ \hline
Methodology & \tabincell{c}{LS channel estimator: \\ $\hat{\mathbf{H}}_{\mathrm{UL}} = f_{\mathrm{LS}}(\widetilde{\mathbf{X}}) = \widetilde{\mathbf{Y}} \widetilde{\mathbf{X}}^H (\widetilde{\mathbf{X}} \widetilde{\mathbf{X}}^H)^{-1}$} & \tabincell{c}{Uplink to downlink channel mapping \cite{Alrabeiah19FDD}: \\ $\hat{\mathbf{H}}_{\mathrm{DL}} = \mathcal{H}(\hat{\mathbf{H}}_{\mathrm{UL}})$} & \tabincell{c}{ZF beamformer: \\ $\mathbf{V} = h_{\mathrm{ZF}}(\hat{\mathbf{H}}_{\mathrm{DL}}) = \gamma_{\mathrm{ZF}} \hat{\mathbf{H}}_{\mathrm{DL}}^{H}(\hat{\mathbf{H}}_{\mathrm{DL}} \hat{\mathbf{H}}_{\mathrm{DL}}^{H})^{-1}$}  \\ \hline
Objective & $\min_{\hat{\mathbf{H}}_{\mathrm{UL}}} ~ \|\hat{\mathbf{H}}_{\mathrm{UL}} \widetilde{\mathbf{X}} - \widetilde{\mathbf{Y}}\|_2^2$ & 
$\min_{\mathcal{H}(\cdot)} ~ \|\mathbf{H}_{\mathrm{DL}} - \hat{\mathbf{H}}_{\mathrm{DL}}\|_2^2$
& $\max_{\mathbf{V}} ~ \sum\limits_{k=1}^{K} \log_2\left(1 +  \frac{\lvert \mathbf{h}_{\mathrm{DL},k}^H \mathbf{v}_k \rvert^2}{ \sum_{j\not=k} \lvert \mathbf{h}_{\mathrm{DL},k}^H \mathbf{v}_j \rvert^2+\sigma_0^2 } \right)$   \\ \hline
Comments & \tabincell{c}{Simple and effective in \\massive MIMO systems} & 
\tabincell{c}{Downlink channel prediction \\ via deep learning techniques \cite{Alrabeiah19FDD}
}
& \tabincell{c}{Widely used in massive MIMO systems \\ due to its asymptotic optimality}   \\ \hline
\end{tabular}
\label{block-by-block}
\end{table*}

\section{System Model and Existing Approaches}\label{sec:sys}

\subsection{System Model}
We consider an FDD massive MIMO system where a BS with $M$ antennas serves $K$ single-antenna users. The uplink and downlink channels between user $k$ and the BS are denoted as $\mathbf{h}_{\mathrm{UL},k} \in \mathbb{C}^{M \times 1}$ and $\mathbf{h}_{\mathrm{DL},k} \in \mathbb{C}^{M \times 1}$, respectively. The multi-path channel model \cite{Alrabeiah19FDD} is adopted and the channel between the BS and user $k$ is given by 
\begin{equation}\label{eq:channel_model}
\mathbf{h}_{i,k}=\sum_{\ell=1}^{L_{p}} \alpha_{i, \ell, k} e^{-j 2\pi f_i \tau_{l,k}} \mathbf{a}_{\mathrm{t}}\left(\theta_{\ell, k}\right), \quad i \in \{\mathrm{UL}, \mathrm{DL}\},
\end{equation}
where $L_{p}$ is the number of propagation paths and $f_i$ denotes the carrier frequency. The delay of the $\ell$-th path is represented by $\tau_{l,k}=\frac{d_{l,k}}{c}$, where $c$ is the speed of light and $d_{l,k}$ denotes the distance of the $\ell$-th path. In addition, $\alpha_{i, \ell, k}$ is the complex gain of the $\ell$-th path, $\theta_{\ell,k}$ is the corresponding angle of departure (AoD), and $\mathbf{a}_\mathrm{t}\left(\cdot\right)$ is the transmit array response vector. For a uniform linear array (ULA) with $M$ antenna elements, the transmit array response vector is given by
\begin{equation}
\mathbf{a}_{\mathrm{t}}(\theta)=\left[1, e^{j \frac{2 \pi}{\lambda} d \sin (\theta)}, \ldots, e^{j \frac{2 \pi}{\lambda} d(M-1) \sin (\theta)}\right]^{T},
\end{equation}
where $\lambda$ is the wavelength and $d$ is the antenna spacing. Since the uplink and downlink channels share the same physical propagation environment, it was shown in \cite{Alrabeiah19FDD} that an intricate mapping from the uplink to the downlink exists and can be approximated by a well-trained MLP.

We consider an uplink training phase, in which each user sends training pilots 
$\widetilde{\mathbf{x}}_k^H \in \mathbb{C}^{1\times L}$ of length $L$.
The received signal $\widetilde{\mathbf{Y}}\in\mathbb{C}^{M\times L}$ at the BS is given by
\begin{equation}\label{eq_rx_pilot}
    \widetilde{\mathbf{Y}} = \sum\limits_{k=1}^{K} \mathbf{h}_{\mathrm{UL},k} \widetilde{\mathbf{x}}_k^H + \widetilde{\mathbf{N}} = \mathbf{H}_{\mathrm{UL}} \widetilde{\mathbf{X}} + \widetilde{\mathbf{N}},
\end{equation}
where $\widetilde{\mathbf{X}} = [\widetilde{\mathbf{x}}_1, \widetilde{\mathbf{x}}_2,\ldots,\widetilde{\mathbf{x}}_K]^H \in \mathbb{C}^{K \times L}$, $\mathbf{H}_{\mathrm{UL}} = [\mathbf{h}_{\mathrm{UL},1}, \mathbf{h}_{\mathrm{UL},2},\ldots,\mathbf{h}_{\mathrm{UL},K}] \in \mathbb{C}^{M \times K}$, and $\widetilde{\mathbf{N}}$ is the additive white Gaussian noise at the BS whose entries are independent and identically distributed (i.i.d.) with zero mean and variance $\sigma_1^2$. The transmitted pilots of each user satisfy the power constraint $\|\widetilde{\mathbf{x}}_k\|_2^2\leq P_{\mathrm{UL}}L$, where $P_{\mathrm{UL}}$ is the maximum uplink transmit power. 

In the downlink data transmission phase, let ${\mathbf{V}}\in \mathbb{C}^{M \times K}$ denote the beamforming matrix at the BS and $\mathbf{s}\in \mathbb{C}^{K\times 1}$ represent the transmit symbol vector. Then, the transmit signal is given by
\begin{equation}
\mathbf{x}=\sum\limits_{k=1}^{K} \mathbf{v}_{k}s_{k}={\mathbf{V}}\mathbf{s}, 
\end{equation}
where $\mathbf{v}_{k}\in \mathbb{C}^{M \times 1}$ is the $k$-th column of ${\mathbf{V}}$ denoting the beamforming vector for user $k$. The transmit power constraint is given by $\operatorname{Tr}(\mathbf{V} \mathbf{V}^H) \leq P_{\mathrm{DL}}$, where $P_\mathrm{DL}$ is the maximum downlink transmit power. The symbol transmitted to user $k$ is denoted by $s_{k}$, where $\mathbb{E}[|s_{k}|^2]=1$.
By assuming a quasi-static flat-fading channel model, the received signal at user $k$ is expressed as
\begin{equation}\label{eq_rx_sig}
y_k =  \mathbf{h}_{\mathrm{DL},k}^H \mathbf{v}_{k} s_k + \sum_{j\not=k} \mathbf{h}_{\mathrm{DL},k}^H \mathbf{v}_{j} s_j + n_k,
\end{equation}
where $n_k \sim \mathcal{CN}(0,\sigma_0^2)$ is the additive white Gaussian noise.
This paper aims to design the beamforming matrix ${\bf{V}}$ at the BS to maximize the system sum-rate
\begin{equation}\label{eq_sumrate}
R=\sum\limits_{k=1}^{K} \log_2\left(1 +  \frac{\lvert \mathbf{h}_{\mathrm{DL},k}^H \mathbf{v}_{k} \rvert^2}{ \sum_{j\not=k} \lvert \mathbf{h}_{\mathrm{DL},k}^H \mathbf{v}_{j} \rvert^2+\sigma_0^2 } \right).
\end{equation}

\subsection{Existing Approaches} \label{existing}
\subsubsection{Beamforming with Explicit Channel Estimation}
The conventional communication design paradigm involves three communication modules, i.e., uplink channel estimation, downlink channel prediction, and downlink beamforming.
Typical techniques employed for these three blocks are shown in Table \ref{block-by-block}, where $\hat{\mathbf{H}}_{\mathrm{UL}}$ denotes the estimated uplink channel, $\hat{\mathbf{H}}_{\mathrm{DL}} = [\hat{\mathbf{h}}_{\mathrm{DL},1}, \hat{\mathbf{h}}_{\mathrm{DL},2},\ldots,\hat{\mathbf{h}}_{\mathrm{DL},K}]^H \in \mathbb{C}^{K \times M}$ denotes the estimated downlink channel, and $\gamma_{\mathrm{ZF}}$ is the scaling constant of ZF beamformer to satisfy the power constraint. 
However, the global optimality is hardly achievable for this block-by-block method. To improve the overall performance, the end-to-end design that fuses different blocks into one is greatly in need.

\subsubsection{Fully Data-Driven End-to-End Design}
For the end-to-end design, the BS extracts the useful information from the received pilots $\widetilde{\mathbf{Y}}$ and then directly design the downlink beamforming $\mathbf{V}$, i.e.,
\begin{equation} \label{functionG}
\mathbf{V} = \mathcal{G}\left( \widetilde{\mathbf{Y}} \right) = \mathcal{G}\left( \mathbf{H}_{\mathrm{UL}} \widetilde{\mathbf{X}} + \widetilde{\mathbf{N}} \right),
\end{equation}
where $\mathcal{G}(\cdot): \mathbb{C}^{M\times L} \rightarrow \mathbb{C}^{M\times K} $ represents the mapping from the received pilots to the downlink beamformer. Note that this method bypasses the explicit channel estimation stage and directly optimizes the beamforming matrix. 
Thus, the sum-rate maximization problem with implicit channel estimation is formulated as
\begin{equation}
\begin{aligned}
\label{main_problem}
\max_{\widetilde{\mathbf{X}},\mathcal{G}(\cdot)} ~ & \sum_{k=1}^{K} \log_2\left(1 +  \frac{\lvert \mathbf{h}_{\mathrm{DL},k}^H \mathbf{v}_k \rvert^2}{ \sum_{j\not=k} \lvert \mathbf{h}_{\mathrm{DL},k}^H \mathbf{v}_j \rvert^2+\sigma_0^2 } \right)\\ 
\text{s.t.} ~ & \mathbf{V} = \mathcal{G}\left( \widetilde{\mathbf{Y}} \right) = \mathcal{G}\left( \mathbf{H}_{\mathrm{UL}} \widetilde{\mathbf{X}} + \widetilde{\mathbf{N}} \right),\\
& \operatorname{Tr}(\mathbf{V} \mathbf{V}^H) \leq P_{\mathrm{DL}},\\
&\|\widetilde{\mathbf{x}}_k\|_2^2\leq P_{\mathrm{UL}} L, ~~\forall k,
\end{aligned}
\end{equation}
in which the uplink training pilots $\widetilde{\mathbf{X}}$ and the beamforming scheme $\mathcal{G}(\cdot)$ are jointly optimized.

In the fully data-driven end-to-end design, the optimal mapping $\mathcal{G}(\cdot)$ can be approximated by neural networks trained with a large amount of data \cite{sohrabi2020deep, Jiang20Implicit}.
However, since fully data-driven methods treat conventional signal processing modules as a black-box without incorporating the domain knowledge, the underlying high-dimensional mapping is difficult to learn in large-scale wireless networks. Therefore, it is difficult for these methods to achieve satisfactory performance when the size of wireless networks scales up. In the next section, we shall propose a scalable end-to-end method that solves \eqref{main_problem} effectively in large-scale wireless systems.

\section{Proposed Neural Calibration Design}
In this section, we introduce the neural calibration-based method for problem in \eqref{main_problem}, where both advantages of the classic linear signal processing methods and data-driven methods are combined.
\subsection{Architecture of the Proposed Neural Calibration}
Although linear signal processing methods are widely adopted due to high computational efficiency, they are only asymptotically optimal for individual communication modules. Therefore, simply cascading them cannot achieve satisfactory performance for the whole system.
To address this problem, we propose a neural calibration-based method that integrates linear signal processing techniques with deep learning. In particular, the linear modules of the classic method are kept in place, and neural networks are adopted to calibrate the inputs to the linear modules. Given the universal approximation property, the added neural networks are able to improve the performance of the end-to-end design.

\begin{figure}[t] 
\centering
\includegraphics[width=0.5\textwidth]{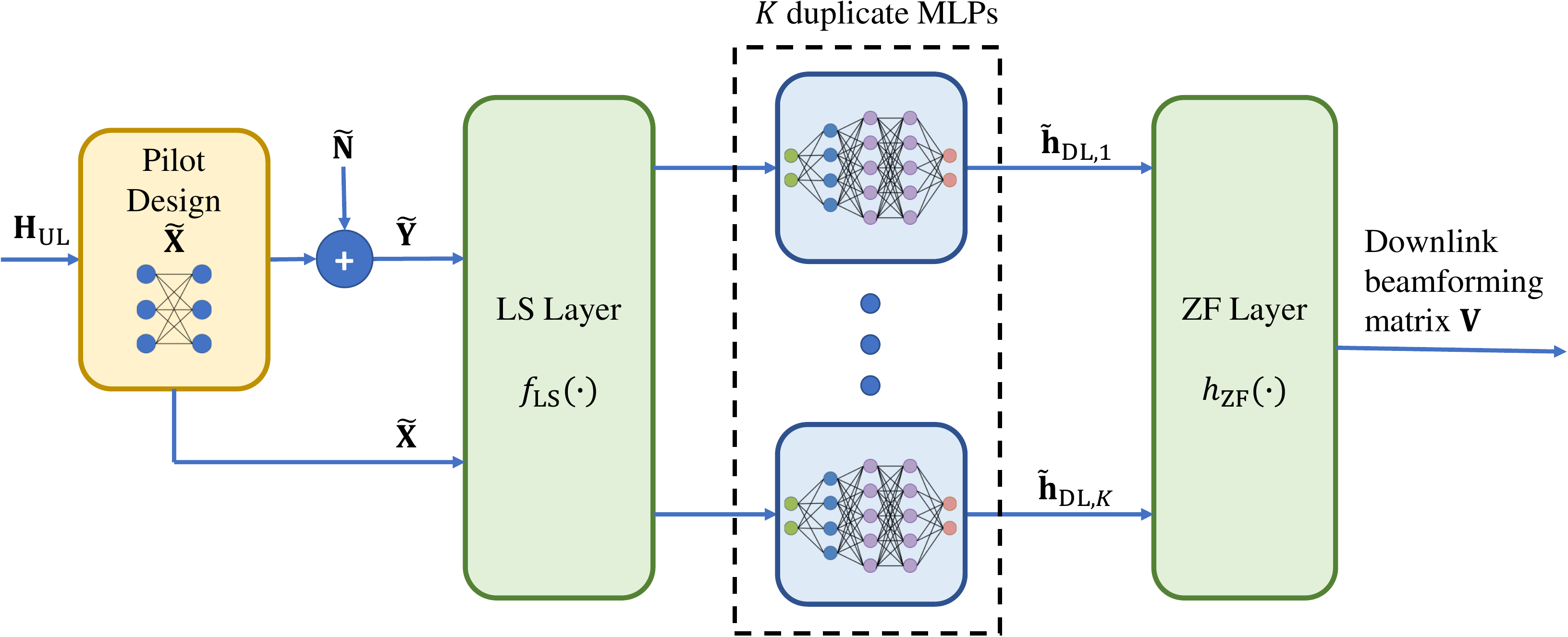} 
\caption{Proposed neural calibration-based end-to-end design architecture in the training stage.} 
\label{architecture} 
\end{figure}

In the following, we first focus on the training stage of the proposed method while the inference stage shall be presented at the end of this section. The overall block diagram of the proposed neural calibration-based end-to-end design in the training stage is shown in Fig. \ref{architecture}. In particular, the backbone of LS estimator, i.e., $f_{\mathrm{LS}}(\cdot)$, is preserved and a pilot design MLP is deployed to improve the overall performance. Besides, a ZF layer, i.e., $h_{\mathrm{ZF}}(\cdot)$, is utilized to perform downlink beamforming, whose input is calibrated by neural networks. Then, the neural calibration-based downlink beamforming optimization with implicit downlink channel estimation is reformulated as
\begin{equation} \label{CM+CaliZF}
\begin{aligned}
\max_{\widetilde{\mathbf{X}}, \mathcal{F} \circ \mathcal{H}(\cdot)} ~& \sum_{k=1}^{K} \log_2\left(1 +  \frac{\lvert \mathbf{h}_{\mathrm{DL},k}^H \mathbf{v}_k \rvert^2}{ \sum_{j\not=k} \lvert \mathbf{h}_{\mathrm{DL},k}^H \mathbf{v}_j \rvert^2+\sigma_0^2 } \right)\\ 
\text{s.t.}  ~ & \mathbf{V} = h_{\mathrm{ZF}}( \mathcal{F} \circ \mathcal{H}(f_{\mathrm{LS}}(\widetilde{\mathbf{X}}) ) ), \\
&\|\widetilde{\mathbf{x}}_k\|_2^2\leq P_{\mathrm{UL}} L, ~~\forall k,
\end{aligned}
\end{equation}
where $\mathcal{F}(\cdot)$ denotes the mapping of the calibration for ZF beamformer, $\mathcal{H}(\cdot)$ denotes the uplink-to-downlink channel mapping \cite{Alrabeiah19FDD}, and $\circ$ represents the composition of two mappings.

\subsection{Uplink Pilot Design}
For LS estimator, conventionally, the optimal pilots are designed according to the mean-square error (MSE) criterion, and one typical choice is the normalized submatrix of the discrete Fourier transform (DFT) matrix \cite{Biguesh06Estimator}. However, the normalized DFT matrix is no longer optimal for the sum-rate maximization problem in \eqref{CM+CaliZF} and thus, a more delicate design for pilot matrix is desired.
In the uplink pilot transmission phase, the users send training pilots $\widetilde{\mathbf{X}}$ and the BS receives $\widetilde{\mathbf{Y}} = \mathbf{H}_{\mathrm{UL}} \widetilde{\mathbf{X}} + \widetilde{\mathbf{N}}$.
Note that a single-layer MLP with a linear activation function can also be regarded as a matrix multiplication operation.
Therefore, to imitate this transmission process and provide a better design of pilots, an MLP is adopted where $\widetilde{\mathbf{X}}$, $\mathbf{H}_{\mathrm{UL}}$, and $\widetilde{\mathbf{Y}}$ are treated as the trainable parameters, input, and output, respectively \cite{sohrabi2020deep}. Specifically, a single-layer MLP with linear activation function and null bias matrix is adopted, followed by an additive zero-mean noise with variance $\sigma_1^2$. Furthermore, to ensure that the designed weight matrix $\widetilde{\mathbf{X}}$ satisfies the uplink power constraint, we adopt an additional weight normalization step in which each $\widetilde{\mathbf{x}}_k$ is normalized to satisfy $\|\widetilde{\mathbf{x}}_k\|^2_2 \le P_{\mathrm{UL}}L$.

\subsection{Neural Calibration for ZF Beamformer}
As introduced in \eqref{CM+CaliZF}, two mappings need to be learned for downlink beamforming. Since the mapping $\mathcal{H}(\cdot)$ from the uplink channel to the downlink channel has been extensively studied in the literature \cite{Alrabeiah19FDD}, we focus on the mapping $\mathcal{F}(\cdot)$ for the calibration of the ZF beamformer and then illustrate how to learn the composite mapping $\mathcal{F}\circ\mathcal{H}(\cdot)$. Before introducing the neural network architecture for calibration, we prove the existence of the mapping $\mathcal{F}(\cdot)$.
\begin{thm}\label{thm1}
(Potential to Calibrate) Define 
\begin{align*}
    &\mathbf{V}_{\mathrm{ZF}} = h_{\mathrm{ZF}}(\mathbf{X}) = \gamma_{\mathrm{ZF}} \mathbf{X}^{H}\left(\mathbf{X} \mathbf{X}^{H}\right)^{-1}, \\
    &R(\mathbf{V}_{\mathrm{ZF}}) = \sum_{k=1}^{K} \log_2\left(1 +  \frac{\lvert \mathbf{h}_{\mathrm{DL},k}^H \mathbf{v}_{\mathrm{ZF},k} \rvert^2}{ \sum_{j\not=k} \lvert \mathbf{h}_{\mathrm{DL},k}^H \mathbf{v}_{\mathrm{ZF},j} \rvert^2+\sigma_0^2 } \right).
\end{align*}
Then, when $\sigma_0^2 \neq 0$, $\forall \mathbf{H}_{\mathrm{DL}}$ and $\forall \epsilon > 0$, there exists an $\widetilde{\mathbf{H}}_{\mathrm{DL}}$ with $\|\mathbf{H}_{\mathrm{DL}} - \widetilde{\mathbf{H}}_{\mathrm{DL}}\| \leq \epsilon$ such that $R\left(h_{\mathrm{ZF}}(\mathbf{H}_{\mathrm{DL}})\right) < R(h_{\mathrm{ZF}}(\widetilde{\mathbf{H}}_{\mathrm{DL}}))$.
\end{thm}
\begin{IEEEproof}
Please refer to the Appendix.
\end{IEEEproof}
Theorem \ref{thm1} indicates that for every channel realization $\mathbf{H}_{\mathrm{DL}}$, there always exists a beamformer $h_{\mathrm{ZF}}(\widetilde{\mathbf{H}}_{\mathrm{DL}})$ other than the conventional ZF beamformer $h_{\mathrm{ZF}}(\mathbf{H}_{\mathrm{DL}})$, which achieves a higher system sum-rate.
Note that even if it exists, it is typically unknown and hard to analytically characterize. 
In this case, the powerful learning capabilities of MLPs can be leveraged to approximate the complicated mapping $\widetilde{\mathbf{H}}_{\mathrm{DL}} = \mathcal{F}(\mathbf{H}_{\mathrm{DL}})$. 

Note that, instead of considering $\mathcal{F}(\cdot)$ and $\mathcal{H}(\cdot)$ as two separate modules and using two different neural networks to approximate them, we directly learn the composite mapping $\mathcal{F} \circ \mathcal{H}(\cdot)$ to improve the overall performance and sidestep the explicit downlink channel reconstruction.
However, the input and output dimensions of the parameterized mapping $\mathcal{F} \circ \mathcal{H}(\cdot)$ are both $2MK$, which is a large number in massive MIMO systems. Employing giant and unstructured neural networks does not incorporate the uniqueness of Problem \eqref{CM+CaliZF} and thus suffers from poor scalability.

\begin{figure}[t] 
\centering
\includegraphics[width=0.22\textwidth]{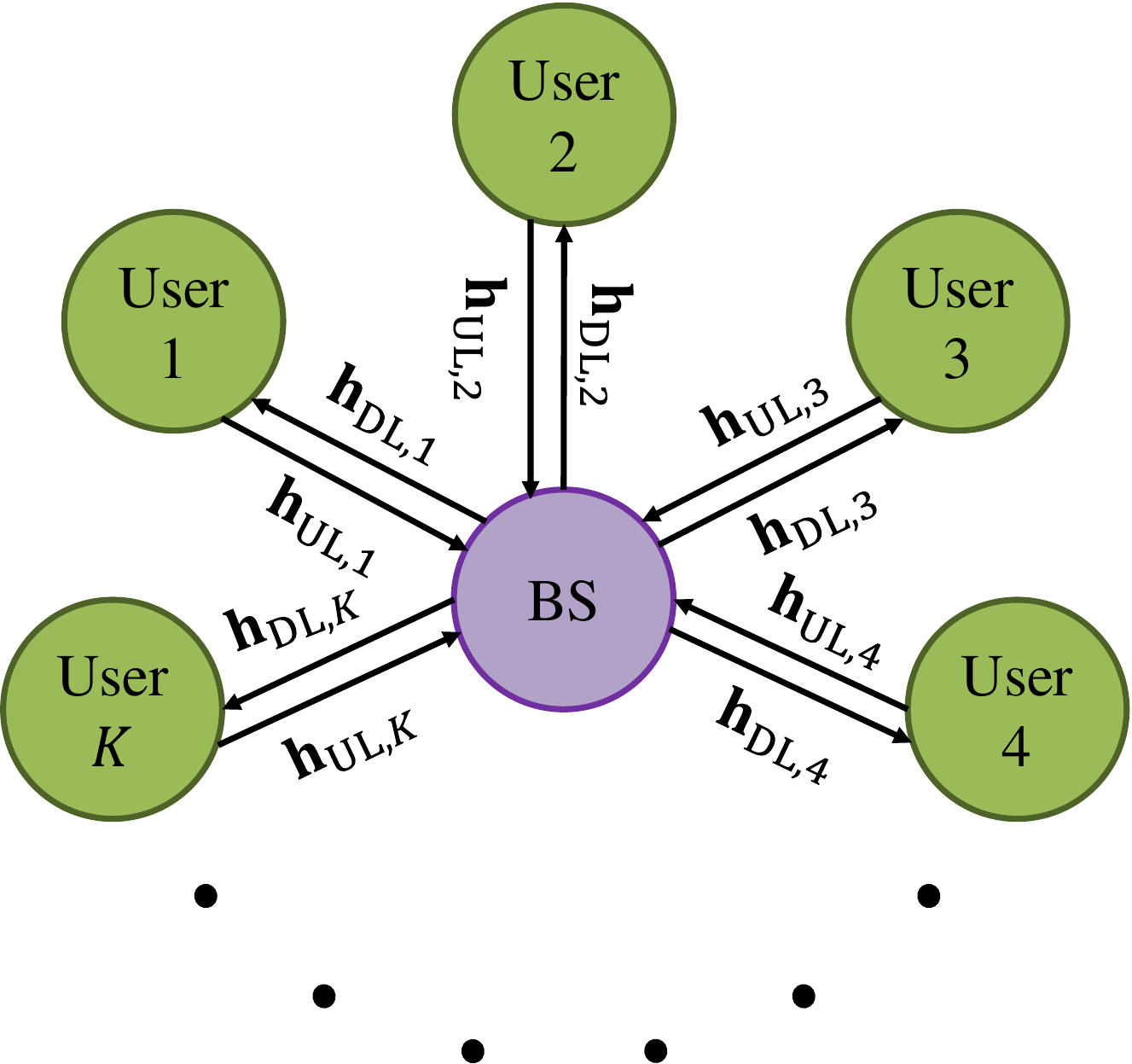}
\caption{Graph modeling of the considered learning problem \eqref{CM+CaliZF}.} 
\label{graph} 
\end{figure}

In this paper, we design a scalable neural network architecture to approximate $\mathcal{F} \circ \mathcal{H}(\cdot)$ by identifying the permutation equivariance property of Problem \eqref{CM+CaliZF}. To this end, we resort to graph theory, which has found abundant successful applications such as radio resource management \cite{Shen21Graph}. 
We model the considered multiuser MIMO network as a directed graph with edge features. In particular, each transmitter or receiver is modeled as one \emph{node} in the graph, and an \emph{edge} is drawn from node $i$ to node $j$ if there is a transmission link from transmitter $i$ to receiver $j$.  Therefore, we obtain a star where the BS is represented by the central node while the $K$ users are represented by $K$ leaves, as shown in Fig. \ref{graph}. 
We denote the $K$ users as node $1, 2, \ldots, K$ and the BS as the $(K+1)$-th node, without loss of generality.
The adjacency feature tensor $\mathbf{A} \in \mathbb{C}^{(K+1) \times(K+1) \times M}$ is given by
\begin{equation}
\mathbf{A}_{(i, j,:)}=\left\{\begin{array}{ll}
\mathbf{0}, & \text { if }\{i, j\} \notin E \\
\text{channel from $i$ to $j$}, & \text { otherwise },
\end{array}\right.
\end{equation}
where $_{(i, j,:)}$ denotes the vector located at the $i$-th row and $j$-th column of a tensor, $\mathbf{0} \in \mathbb{C}^{M \times 1}$ is a zero vector, and $E$ denotes the set of edges. Then, \eqref{CM+CaliZF} can be rewritten as an optimization problem over a graph, given by
\begin{equation} \label{perm_CM+CaliZF}
\begin{aligned}
\max_{\widetilde{\mathbf{X}}, \mathcal{F} \circ \mathcal{H}(\cdot)} ~& \sum_{k=1}^{K} \log_2\left(1 +  \frac{\lvert \mathbf{A}_{(K+1, k,:)}^H \mathbf{v}_k \rvert^2}{ \sum_{j\not=k} \lvert \mathbf{A}_{(K+1, k,:)}^H \mathbf{v}_j \rvert^2+\sigma_0^2 } \right)\\ 
\text{s.t.}  ~ & \mathbf{V} = h_{\mathrm{ZF}}( \mathcal{F}\circ \mathcal{H}(f_{\mathrm{LS}}(\widetilde{\mathbf{X}}) ) ),\\
~ & \|\widetilde{\mathbf{x}}_k\|_2^2 \leq P_{\mathrm{UL}} L, ~~\forall k,
\end{aligned}
\end{equation}
where $f_{\mathrm{LS}}(\widetilde{\mathbf{X}}) = \widetilde{\mathbf{Y}} \widetilde{\mathbf{X}}^H (\widetilde{\mathbf{X}} \widetilde{\mathbf{X}}^H)^{-1} = (\sum\nolimits_{k=1}^{K} \mathbf{A}_{(k, K+1,:)} \widetilde{\mathbf{x}}_k^H + \widetilde{\mathbf{N}})\widetilde{\mathbf{X}}^H (\widetilde{\mathbf{X}} \widetilde{\mathbf{X}}^H)^{-1}$. It is shown in \cite[Proposition 3]{Shen21Graph} that the permutation equivariant property is universal for optimization problems over a graph. Therefore, Problem \eqref{perm_CM+CaliZF} enjoys a permutation equivariance property with respect to $\mathbf{V}$ and $\mathbf{A}$.

Since permuting $\mathbf{V}$ and $\mathbf{A}$ simultaneously is simply a reordering of the nodes, the equivariance property suggests that it is the elements in $\mathbf{A}$ rather than the ordering of those elements that count when maximizing the sum-rate. This implies that all the edges with the same end node (the channels in Fig. \ref{graph}) are homogeneous and allows us to share trainable weights among different users. Inspired by this important observation, we develop $K$ duplicate MLPs that share the common trainable parameters for $K$ users to approximate the mapping $\mathcal{F} \circ \mathcal{H}(\cdot)$. The input and output dimension of each MLP is then reduced to $2M$, which is independent of the number of users. Moreover, instead of utilizing a centralized MLP or $K$ distinct MLPs, the proposed architecture leads to only one shared lightweight learning model that enables faster training and inference.
It is also important to note that the proposed neural calibration architecture can be interpreted as a one-layer GNN, for which the permutation equivariance property has been proved in \cite[Proposition 5]{Shen21Graph}.

\emph{Remark:} Since the conventional linear estimator and linear beamformer are asymptotically optimal for channel estimation and beamforming, respectively, they are scalable for large-scale wireless networks in terms of the achievable performance. In this way, by exploiting the domain knowledge inherent in these time-efficient methods and incorporating the intrinsic equivariance property introduced by the wireless network topology, the proposed neural calibration-based end-to-end design is more scalable than the fully data-driven end-to-end method, which shall also be verified via simulations in the next section.

Note that the accurate CSI is only needed in the training stage. After training, the pilot matrix $\widetilde{\mathbf{X}}$ and the trainable parameters in $K$ duplicate MLPs are fixed. In the inference stage, the received pilots $\widetilde{\mathbf{Y}}$ and pilot matrix $\widetilde{\mathbf{X}}$ are fed into the LS layer and the downlink beamforming matrix is obtained consequently.

\section{Simulation Results} \label{simulation}
In this section, we demonstrate the performance of the proposed neural calibration-based end-to-end design in FDD massive MIMO systems. 

\subsection{Simulation Setup}
In the experiments, we consider the indoor massive MIMO scenario ``I1'' that is provided in the DeepMIMO dataset \cite{Alkhateeb2019} and constructed based on the 3D ray tracing simulator. 
Two operating frequencies, i.e., 2.4 GHz and 2.5 GHz, are employed as the uplink and downlink carrier frequencies, respectively.
We consider the ULA transmit antennas at the BS with the antenna spacing set at half wavelength and the number of propagation paths set to 5. 
The uplink pilot length is the same as the number of users and the noise power is set to $\sigma_0^2=\sigma_1^2=-85$ dBm in the simulations.

The user-wise shared MLP adopted in the proposed method has 4 fully-connected layers with 512, 2048, 2048, and $2M$ neurons in each layer, respectively. The objective function in \eqref{CM+CaliZF} is used as the unsupervised loss. We train the neural network for 200 epochs using the Adam optimizer with a mini-batch size of 1024 and a learning rate of 0.001. There are in total 204,800 training samples and 1000 test samples. After each dense layer, the batch normalization layer is leveraged to accelerate convergence and the rectified linear unit (ReLU) is utilized as the activation function in the hidden layers.

\subsection{Performance Comparison}
To illustrate the effectiveness of the proposed neural calibration end-to-end design, we adopt four benchmarks for comparisons:
\begin{itemize}
\item \textbf{WMMSE Perfect CSIT}: Assuming perfect downlink CSI at the transmitter (CSIT), the conventional iterative WMMSE method in \cite{Shi11WMMSE} is used for beamforming. This baseline serves as a performance upper bound with a high computational complexity.
\item \textbf{ZF Perfect CSIT}: The ZF beamforming solution is employed given perfect CSIT.
\item \textbf{Fully Data-driven End-to-End}: The black-box CNN method in \cite{Hu21Unfold} is modified to map the received uplink pilot to the downlink precoding matrix.
\item \textbf{LS + Channel Mapping + ZF}: Conventional block-by-block methodologies in Table \ref{block-by-block} are utilized.
\end{itemize}

\begin{figure}[t] 
\centering
\includegraphics[height=5.6cm]{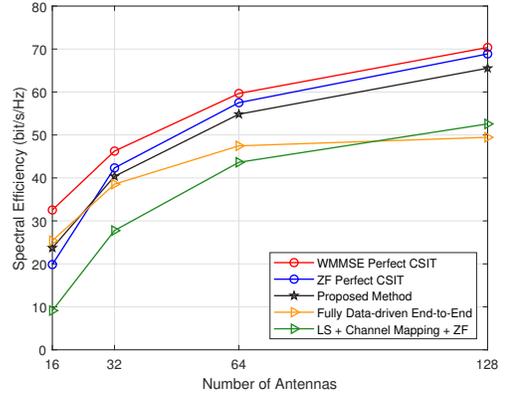} 
\caption{Spectral efficiency achieved by different methods when $K=8$, $P_{\mathrm{UL}}=-10$ dBm, and $P_{\mathrm{DL}}=5$ dBm.} 
\label{figure1} 
\end{figure}

\begin{figure}[t] 
\centering
\includegraphics[height=5.6cm]{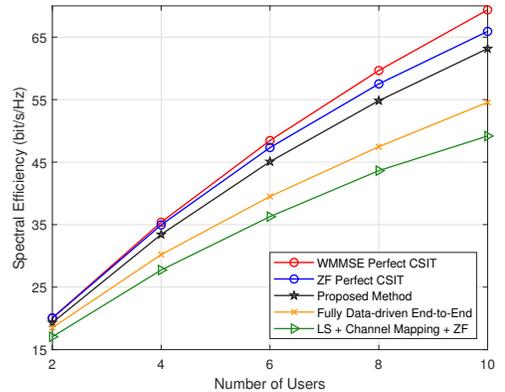} 
\caption{Spectral efficiency achieved by different methods when $M=64$, $P_{\mathrm{UL}}=-10$ dBm, and $P_{\mathrm{DL}}=5$ dBm.} 
\label{figure2} 
\end{figure}
Fig. \ref{figure1} plots the spectral efficiency achieved by the proposed scheme and the four baseline methods versus the number of antennas. It can be observed that the proposed neural calibration-based design significantly outperforms the conventional block-by-block method for all investigated values of antenna size.
Besides, the proposed method even outperforms ZF with perfect CSIT when the number of antennas $M$ is small. This shows that, by calibrating the input of the ZF beamformer using neural networks, the system performance in the small-scale antenna regime can be effectively improved. Furthermore, it is demonstrated in Fig. \ref{figure1} that the performance of our proposed method captures the trend of the two baseline beamformers with perfect CSIT. In contrast, the fully data-driven end-to-end method suffers from a significant performance degradation when $M$ increases, and its performance gain over the block-by-block method vanishes when $M = 128$. This clearly indicates that, by integrating domain knowledge into the end-to-end design, our approach achieves a higher scalability compared to the fully data-driven end-to-end method \cite{Hu21Unfold}.

In Fig. \ref{figure2}, we demonstrate the system spectral efficiency versus the number of users. As can be observed in Fig. \ref{figure2}, while the two baseline methods without perfect CSIT entail a prominent performance loss when the number of users increases, our proposed neural calibration-based method only experiences little performance loss.
Specifically, the proposed beamforming method can achieve $91.1\%$ of the spectral efficiency achieved by the WMMSE with perfect CSIT when $K=10$, while the fully data-driven end-to-end method only achieves $78.7\%$ in this case. This verifies the superiority of the proposed neural calibration design in terms of both the spectral efficiency and scalability in wireless networks where users are densely distributed.

\begin{figure}[t] 
\centering
\includegraphics[height=5.6cm]{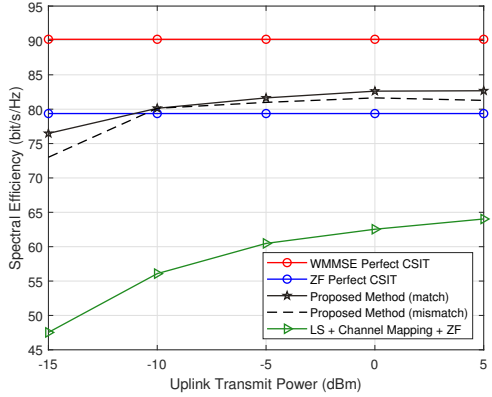} 
\caption{Spectral efficiency achieved by different values of uplink transmit power when $M=64$, $K=16$, and $P_{\rm{DL}}=5$ dBm.} 
\label{figure3} 
\end{figure}
We also test the robustness of our method against mismatch in terms of the uplink SNR. In Fig. \ref{figure3}, the ``Proposed Method (mismatch)" indicates the case where the learning model is trained when $P_{\rm{UL}}=-10$ dBm but tested for different values of uplink transmit power. 
It is shown that there is little performance loss in terms of the spectral efficiency when mismatch exists for the uplink transmit power, which confirms the robustness of the proposed neural calibration end-to-end design. Besides, the spectral efficiency achieved by both the matched and mismatched neural calibrations is significantly higher than that of the block-by-block design over the whole SNR regime, which indicates the superiority of neural calibration-based resource allocation with implicit channel estimation. Since the fully data-driven end-to-end benchmark is not scalable for dense wireless networks, its performance is not shown here.

\section{Conclusions}
In this paper, we developed a neural calibration-based method for downlink beamforming in FDD massive MIMO systems with implicit channel estimation. In contrast to existing fully data-driven methods that completely replace conventional communication modules with DNNs, we provided a novel way to amalgamate domain knowledge with deep learning. Notable advantages of the proposed method include higher spectral efficiency, improved scalability, as well as, robustness against system parameters. Simulation results clearly demonstrated that the proposed neural calibration-based end-to-end design achieves an excellent performance in large-scale FDD systems without explicit downlink CSI.

\appendix
\section{}\label{appA}
\begin{IEEEproof}
The gradient of the sum-rate function with respect to $\mathbf{V}$ is ${\nabla _{\mathbf{V}}}R = {\mathbf{H}}_{\mathrm{DL}}^H{\mathbf{B}}$ where $\mathbf{B} \in \mathbb{C}^{K \times K}$ and its entries are given by
\begin{equation}
\begin{aligned}
{b_{kk}} = ~& \frac{{\mathbf{h}_{\mathrm{DL},k}^H{{\mathbf{v}}_k}}}{{\sum\limits_{i = 1}^K {|\mathbf{h}_{\mathrm{DL},k}^H{{\mathbf{v}}_i}{|^2} + {\sigma_0 ^2}} }}, \\
{b_{jk}} = ~& \frac{{ - |\mathbf{h}_{\mathrm{DL},j}^H{{\mathbf{v}}_j}{|^2}\mathbf{h}_{\mathrm{DL},j}^H{{\mathbf{v}}_k}}}{{(\sum\limits_{i = 1}^K {|\mathbf{h}_{\mathrm{DL},j}^H{{\mathbf{v}}_i}{|^2} + {\sigma_0 ^2}} )(\sum\limits_{i \ne j} {|\mathbf{h}_{\mathrm{DL},j}^H{{\mathbf{v}}_i}{|^2} + {\sigma_0 ^2}} )}}, j \neq k.
\end{aligned}
\end{equation}
Since $\mathbf{V}_{\mathrm{ZF}} = h_{\mathrm{ZF}}(\mathbf{X})$, the gradient of the sum-rate function with respect to the input of ZF, i.e., $\mathbf{X}$, is given by
\begin{equation}
\begin{aligned}
{\nabla _{\mathbf{X}}}R = {\gamma_{\mathrm{ZF}}}[ ~& {(\mathbf{X}{\mathbf{X}^H})^{ - 1}} \mathbf{B}^{H} \mathbf{H}_{\mathrm{DL}}  - \mathbf{X}^ {'} \mathbf{H}_{\mathrm{DL}}^{H} \mathbf{B} \mathbf{X}^ {'} \\
~& - {(\mathbf{X}{\mathbf{X}^H})^{ - 1}} \mathbf{B}^{H} \mathbf{H}_{\mathrm{DL}} {\mathbf{X}^{H}} \mathbf{X}^ {'} ],
\end{aligned}
\end{equation}
where $\mathbf{X}^ {'} = {({\mathbf{X}}{{\mathbf{X}}^H})^{ - 1}} {\mathbf{X}}$.
It can be observed that when $\sigma_0^2 \neq 0$, the gradient of the sum-rate function with respect to $\mathbf{X}$ is not zero at $\mathbf{X}=\mathbf{H}_{\mathrm{DL}}$.
Therefore, when $\sigma_0^2 \neq 0$, $\forall \mathbf{H}_{\mathrm{DL}}$ and $\forall \epsilon > 0$, there exists an $\widetilde{\mathbf{H}}_{\mathrm{DL}}$ with $\|\mathbf{H}_{\mathrm{DL}} - \widetilde{\mathbf{H}}_{\mathrm{DL}}\| \leq \epsilon$ such that $R\left(h_{\mathrm{ZF}}(\mathbf{H}_{\mathrm{DL}})\right) < R(h_{\mathrm{ZF}}(\widetilde{\mathbf{H}}_{\mathrm{DL}}))$. 
\end{IEEEproof}

\bibliographystyle{IEEEtran}
\bibliography{IEEEabrv,FDD_V14}

\end{document}